\documentclass[10pt,twocolumn,letterpaper]{article}

\usepackage{cvpr}
\usepackage{times}
\usepackage{epsfig}
\usepackage{graphicx}
\usepackage{amsmath}
\usepackage{amssymb}
\usepackage{bm}
\usepackage{xcolor}
\usepackage{soul}
\usepackage[utf8]{inputenc}
\usepackage[small]{caption}
\usepackage{multirow}
\usepackage{dsfont}
\usepackage{subfigure}
\usepackage{amsfonts}
\usepackage{epstopdf}
\usepackage{caption}
\usepackage{wrapfig}
\usepackage{color}
\usepackage[ruled,vlined,linesnumbered]{algorithm2e} 
\usepackage{autobreak} 
\usepackage{multirow} 

\cvprfinalcopy 

\def\cvprPaperID{****} 

\begin{document}

\title{Scan-flood Fill(SCAFF): an Efficient Automatic Precise Region Filling Algorithm for Complicated Regions}

\author{
   Yixuan He$^{1,2}$ $\qquad$ Tianyi Hu$^{1,2}$\qquad \qquad \qquad Delu Zeng$^1$\thanks{Corresponding author.}\\
   \tt\small{\{Y.He-48, T.Hu-9\}@sms.ed.ac.uk} \qquad \qquad \qquad \tt\small{dlzeng@scut.edu.cn}\\
   $^1$ South China University of Technology;
   $^2$ The University of Edinburgh.
}
\maketitle

\pagestyle{empty}
\thispagestyle{empty}

\begin{abstract}
 Recently, instant level labeling for supervised machine learning requires a considerable number of filled masks. In this paper, we propose an efficient automatic region filling algorithm for complicated regions. Distinguishing between adjacent connected regions, the Main Filling Process scans through all pixels and fills all the pixels except boundary ones with either exterior or interior label color. In this way, we succeed in classifying all the pixels inside the region except boundary ones in the given image to form two groups: a background group and a mask group. We then set all exterior label pixels to background color, and interior label pixels to mask color. With this algorithm, we are able to generate output masks precisely and efficiently even for complicated regions as long as  boundary pixels are given. Experimental results show that the proposed algorithm can generate precise masks that allow for various machine learning tasks such as supervised training. This algorithm can effectively handle multiple regions, complicated `holes' and regions whose boundaries touch the image border. By testing the algorithm on both toy and practical images, we show that the performance of Scan-flood Fill(SCAFF) has achieved favorable results.
\end{abstract}

\section{Introduction}

\begin{figure}[htbp]
    \begin{center}
        \includegraphics[width=1\columnwidth]{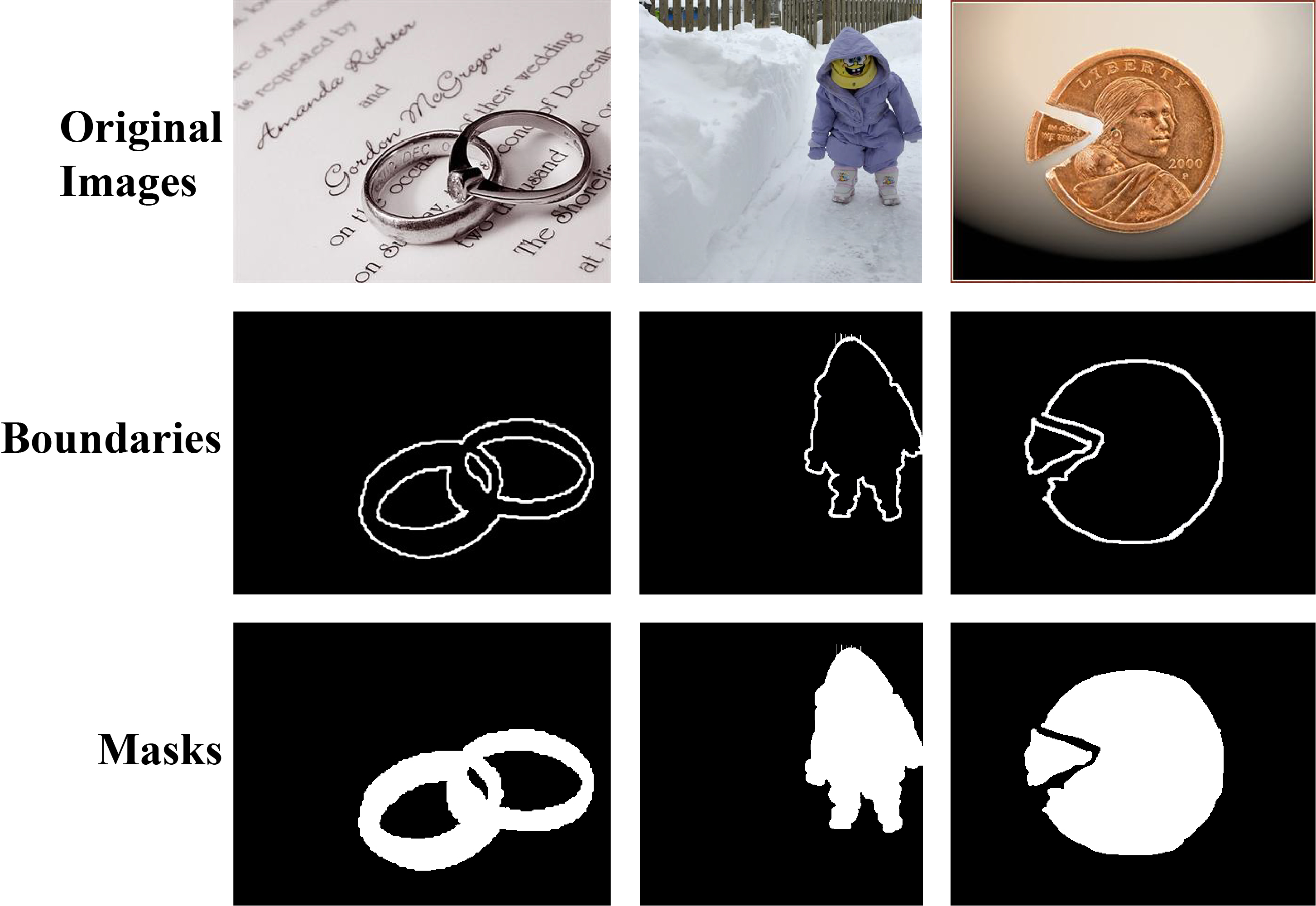}
    \end{center}
    \caption{A practical example for region filling in generating masks for supervised machine learning.}
    \label{fig:practical_example}
\end{figure}

Nowadays, big datasets are widely used in training models for supervised machine learning in many areas, such as salient object segmentation \cite{ren2014region, wang2015deep, 23} and lung nodule detection \cite{Mcnitt-GrayMichaelF.2007TLID}. Although bounding box method can be used to quickly generate training labels \cite{ZhongBineng2016CDBN,HeK.2018MR}, instant-level based learning such as segmentation, recognition and detection usually depend on precise masks \cite{cai2018menet,ren2014region, wang2015deep, 23}. To compress the size of the datasets, some of them only annotate the boundary pixels of instances instead of providing all mask pixels. For instance, the LIDC-IDRI dataset \cite{Mcnitt-GrayMichaelF.2007TLID} widely used for lung nodule detection only provides .XML files that contain the edges of lung nodules. However, in order to achieve better performance on neural networks, especially convolution-based neural networks \cite{ronneberger2015u, liu2016dhsnet}, it is usually important to generate masks, for training. Correspondingly, filling masks from the annotated boundaries is usually a compulsory process in generating ground truth labels for instances. Figure \ref{fig:practical_example} provides an example of how region filling can be used to generate masks for salient object segmentation, given a label image with merely boundary pixels.

In general, region filling refers to an approach to fill some bounded regions with given colors. Based on the domain of the graphics they operate in, region filling algorithms can be classified into raster filling and vector filling \cite{CodreaMariusC.2005NAaf}. In this paper, we focus on raster filling algorithms, \ie on filling regions in raster graphics. This method means filling connected components (defined mostly by 4-connected or 8-connected regions) with required colors. As an important algorithm of Computer Graphics, region filling has various applications in areas such as Computer Aided Design, Realistic graphics, Geographic Information System and Image processing. \cite{Duo-LeF.2011Anfr}

 Generally, region filling algorithms can be categorized as seed filling and edge filling (or scan-line filling) \cite{PavlidisTheodosios1981Afga}. However, current algorithms have several shortcomings or potential problems that need to be considered.
 
 Firstly, most of the filling algorithms cannot run automatically. Many seed filling algorithms, such as ordinary boundary fill and flood fill, require at least one known interior pixel within the boundary. Therefore, they are dependent on operator-provided seeds. For complicated objects of interests, multiple seeds may be required and they can be quite hard to be automatically detected\cite{CodreaMariusC.2005NAaf}.
 
 Secondly, some of the filling algorithms cannot deal with complex boundaries. Scan-fill algorithm is capable of filling a polygon's boundaries \cite{Diehl;MichaelR1997Slgf}, but may mostly fail to deal with arbitrary boundaries. 

Thirdly, it is possible that some of the interior regions' boundary pixels can lie on the border of the image. Hence, it should not be assumed that pixels on the border of an image are all background pixels. The mutual exterior of inner shapes can also be disconnected.

Also, some objects have `holes' inside. If we simply consider filling connected regions that are not connected to the border of the whole image, we may fail to obtain the precise result.

To overcome these drawbacks, and to reduce the tedious or time-consuming manual work when generating ground truth masks for machine learning, we devise a novel region filling algorithm called {\bf Scan-flood Fill(SCAFF)} algorithm. To show that the algorithm resolves all problems mentioned here, we focus on scenarios described in Table \ref{table:case}.




\begin{table}[h!]
    \centering
        \caption{Scenarios for images to be considered in region filling.}
    \begin{tabular}{|c|c|c|c|}
    \hline
        Case & \#Objects        & Touch Border & `Holes' in Object \\ \hline
        1    & 1               & No     & No              \\ \hline
        2    & 1               & No     & Yes             \\ \hline
        3    & 1               & Yes    & No              \\ \hline
        4    & 1               & Yes    & Yes             \\ \hline
        5    & \textgreater{}1 & No     & No              \\ \hline
        6    & \textgreater{}1 & No     & Yes             \\ \hline
        7    & \textgreater{}1 & Yes    & No              \\ \hline
        8    & \textgreater{}1 & Yes    & Yes             \\ \hline
    \end{tabular}

    \label{table:case}
\end{table}

In summary, the main contributions of our work include: 1) We propose an efficient automatic precise region filling algorithm for complicated regions, robustly dealing with cases listed in Table \ref{table:case}. 2) Scan-flood Fill(SCAFF) algorithm frees operators from having to provide starting seeds for region filling. It can detect seeds automatically. 3) The basic version of our proposed algorithm, EFCI, can deal with regions without `holes' effectively.

\section{Related Works}
One of the most common region filling algorithms is seed filling algorithm, or flood filling algorithm, which is usually based on the notion of 4-connectivity or 8-connectivity. Mathematically, let $P = (x,y)$ denote the coordinate of a pixel, then its {\bf 4-Connected region} $\mathbf{C_4}(P)$ in the bitmap is defined as 
\begin{equation}
    \mathbf{C_4}(P) = \left \{(x,y-1),(x,y+1), (x-1,y),(x+1,y)\right\},
\end{equation}
and its {\bf 8-Connected region} $\mathbf{C_8(P)}$ in the bitmap is defined as 

\begin{align}
    \begin{autobreak}
        \mathbf{C_8}(P) = \mathbf{C_4}(P)  \cup
         \{ (x-1,y-1),(x+1,y+1),
        (x-1,y+1),(x+1,y-1) \},
    \end{autobreak}
\end{align}

\begin{figure}[htbp]
    \begin{center}
        \includegraphics[width=1\columnwidth]{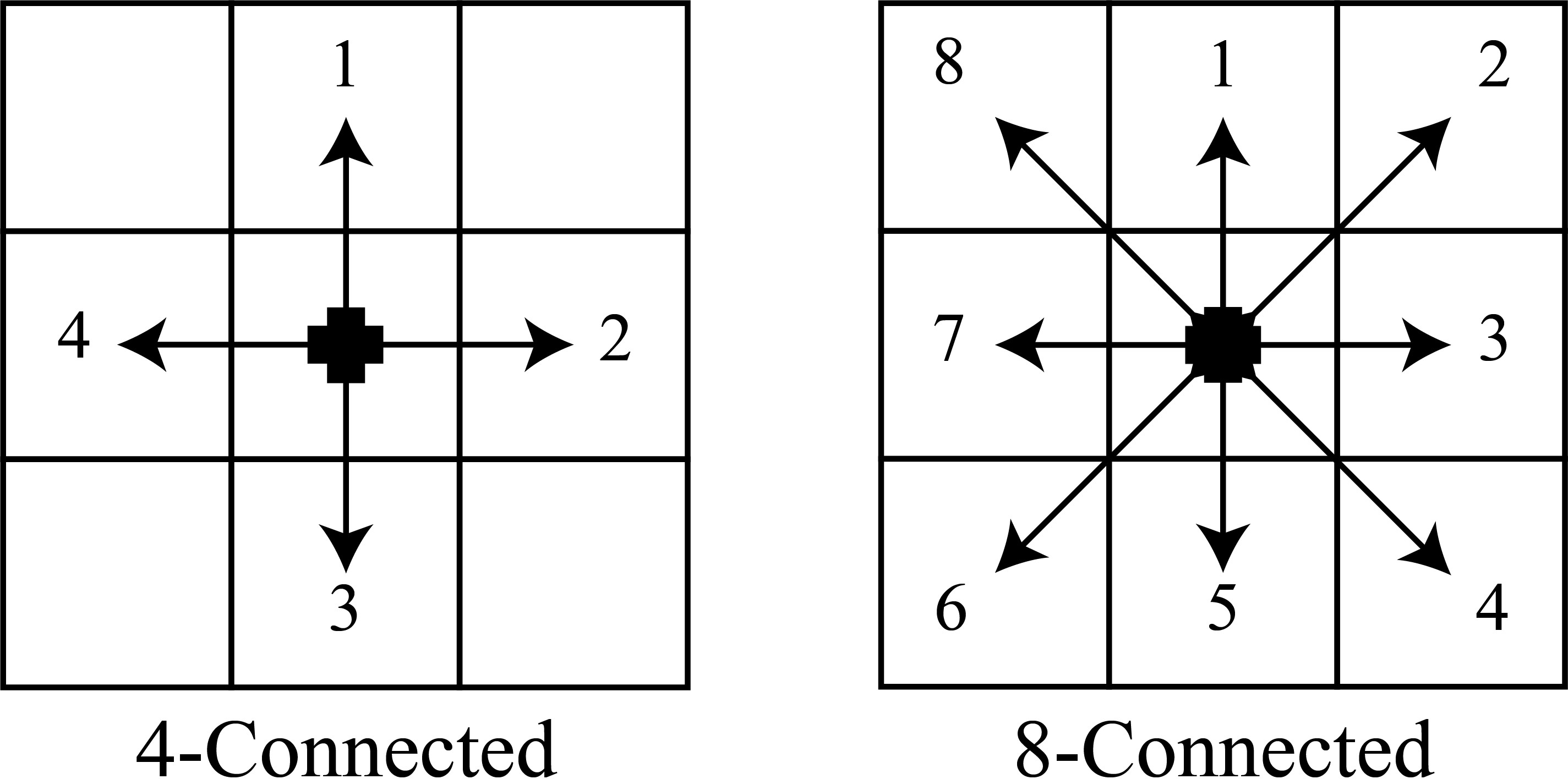}
    \end{center}
    \caption{Concept of connectivity. The pixel in the middle of the box is the pixel of interest, and its 4-connected of 8-connected region consists of all pixels that can be reached by arrows starting from it.}
    \label{fig:connect}
\end{figure}

A visual illustration is given in Figure \ref{fig:connect}.

Seed filling algorithm starts from a known pixel in a closed area, and recursively finds all the pixels within the connected region of the known pixel. To reach all pixels in each region (normally a 4-connected or 8-connected region as defined above), a stack may be required, and pixels within the region will be visited recursively \cite{XungenLi2010Nrfa}.Depth-first search (DFS) algorithm \cite{TarjanRobert1972DSaL} can be used to fill a certain region from a starting seed. It is efficient in time, but costly in space. Therefore, `span' filling methods were proposed to reduce the considerable stack required for searching \cite{foley1996computer}. In 1994, Henrich \cite{HenrichDominik1994Srfi} analyzed several algorithms that saved memory regardless of speed. After that, Yanovsky \etal \cite{YanovskyVladimirM.2007ALCA} proposed a linear-time constant-space algorithm for the boundary fill problem. However, this algorithm cannot lower the efficiency overhead of revisiting the same nodes. Later, Duo-Le and Ming\cite{Duo-LeF.2011Anfr} suggested a `Marking Method' to tackle the problem of unnecessary pixel revisiting. 

Edge filling is another popular approach in region filling, primarily for polygon filling \cite{Diehl;MichaelR1997Slgf, anitha2013efficient}. This algorithm relies on the detection of the spans of the scan line lying inside the polygon, using the odd-parity rule \cite{GonzalezRafaelC.2018Dip}. However,filling by parity may fail when we obtain the number of line segments incorrectly \cite{RenMingwu2005Anaf}.Although scan line filling algorithms are initially designed for polygons, some authors generalize them to wider applications. For example, Cai \cite{CaiZuguang1988Robi} took into account the connectivity of pixels in the same region and also made use of a scanning line. Wherever the closed area intersects the scanning line, the scanning line seed filling algorithm takes merely one seed pixel and starts to fill from it to the left and right directions. Nevertheless, repeated judging of the pixel colors and unnecessary backtracking operations lower the efficiency of this algorithm \cite{CodreaMariusC.2005NAaf}. Vučković \etal\cite{VuckovicV.2017GNis} introduced a generalized iterative scanline fill algorithm useful in real-time applications. In this paper, we adopt some ideas of scan line filling and apply them to label different connected regions. 

\begin{figure*}[t]
    \begin{center}
        \includegraphics[width=1\textwidth]{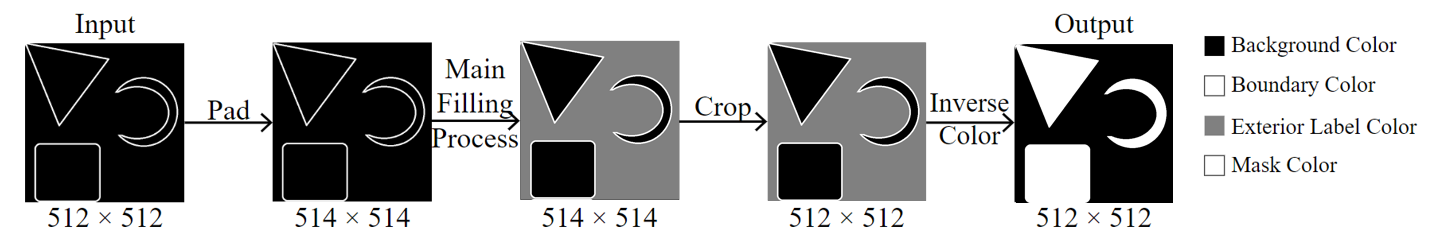}
    \end{center}
    \caption{Evolution of an image during the the implementation of EFCI. The flow chart illustrates how EFCI acts on an image: first pads the image, then fills the exterior with exterior label color. After cropping, the algorithm `inverses' colors to resulting ones.}
    \label{fig:ProcedureBasic}
\end{figure*}

It is usually difficult to set seeds automatically, and searching for all interior pixels may be time and memory
consuming\cite{CodreaMariusC.2005NAaf,RenMingwu2005Anaf}. To automatically find the starting seed, Khayal \etal proposed a modified algorithm for seed filling \cite{KhayalM.S.H.2011Mnaf}. However, this method needs to discover all contour pixels, and compute their angles. In our algorithm, contour computation is omitted, but we use region labels instead.

Making use of {\bf Connected Component Labeling (CCL)}, which is a fundamental operation in image processing\cite{FannySpagnolo2018AECC}, Miao \etal\cite{Miao} proposed a regional filling algorithm based on connected region labeling. However, they did not distinguish between different `interior' regions, but filled all of them in the same color. Our basic algorithm adopts this idea, and then develops a more precise algorithm using similar ideas to CCL.

\section{Proposed Algorithm}

In this paper, we propose an automatic(with respect to setting starting seeds for filling) region filling algorithm, built on a basic version. The proposed algorithm is to fill arbitrary regions in a given image, based on the observation that the exterior of a region has to be connected to the border of the whole image. Neither of the basic version and the improved version of our algorithm requires a seed to be given by the operator at the beginning. The basic version of our algorithm can fill arbitrary regions without interior `holes' precisely, regardless of how many regions there are in the image, and can also handle the case where some regions have some of their boundary pixels on the border of the whole image. The proposed algorithm(\ie the improved version built on the basic version) takes into account potential `holes' inside a region. Those versions of the algorithm are described as follows.

\subsection{Basic Version: Exterior-Fill and Color Inversion (EFCI) }

Our method makes use of the property of the exterior mentioned above. The algorithm starts from an image with only boundary pixels in boundary color and the rest in background color. Padding in background color is first added. Then, the mutual exterior of the regions is filled with a temporary exterior label color. In version~\ref{algo:Basic} of the algorithm (where `floodfill' refers to flood filling algorithm based on the color of the seed, such as OpenCV floodfill \cite{bradski2008learning,howse2013opencv}), the color of the mutual exterior is set to background color, while the interior is filled with the desired mask color. We call the latter process {\bf Crop-and-`Inverse'} process. `Inverse' here means to change colors. That is why this basic version of the algorithm can be called {\bf Exterior-Fill and Color Inversion (EFCI).} An example of how the image changes are given in Figure \ref{fig:ProcedureBasic}.

\begin{algorithm}[h]
    \KwIn{img}
    \KwOut{resultImg}
    padImg $\gets$ pad img with background color\;
    padImg $\gets$ floodfill(padImg,seed = (0,0)) with exterior label color\;
    croppedImg $\gets$ crop padImg to original size, delete padding\;
    resultImg $\gets$ croppedImg\;
    \For{x in range(resultImg.height) }{   \tcp*{color inversion}
        \For{y in range(resultImg.width)}{
            \If{resultImg[x][y] == background color}{
                resultImg[x][y] $\gets$ mask color\;
            }
            \If{resultImg[x][y] == exterior label color}{
                resultImg[x][y] $\gets$ background color\;
            }
        }
     }
    \Return resultImg
    \caption{Exterior-Fill and Color Inversion(img)}
    \label{algo:Basic}
\end{algorithm}

In order to start from a pixel in background color, the padding is needed, since the origin of the original image may be a boundary pixel. Besides, multiple regions can be filled within one implementation, without requiring initial seeds inside them to start from. However, since EFCI can only capture the mutual exterior region, which is connected to the origin in the padded image, it cannot handle the case where there are `holes' inside regions that should not be filled.

\subsection{Filling Complicated Regions with Interior `Holes'}
\begin{figure}[htbp]
    \centering
    \subfigure[Input Image]{%
    \includegraphics[width=3.75cm]{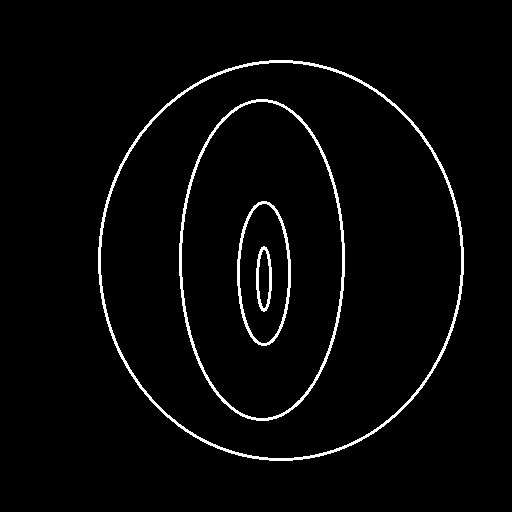}
    \label{fig:HoleInput}
    }
    \quad
    \subfigure[Image Region]{%
    \includegraphics[width=3.75cm]{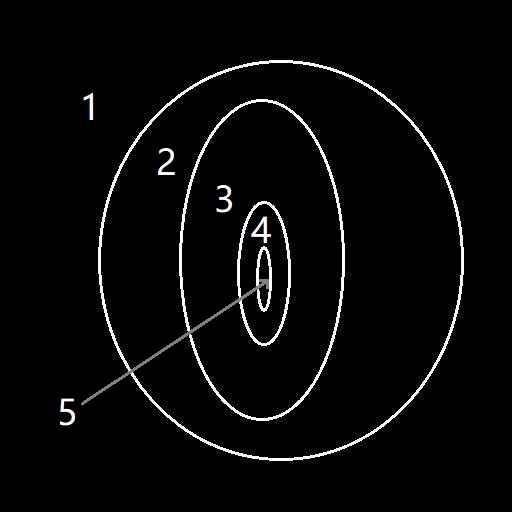}
    \label{fig:HoleRegion}
    }
    \quad
    \subfigure[Incorrect Fill]{%
    \includegraphics[width=3.75cm]{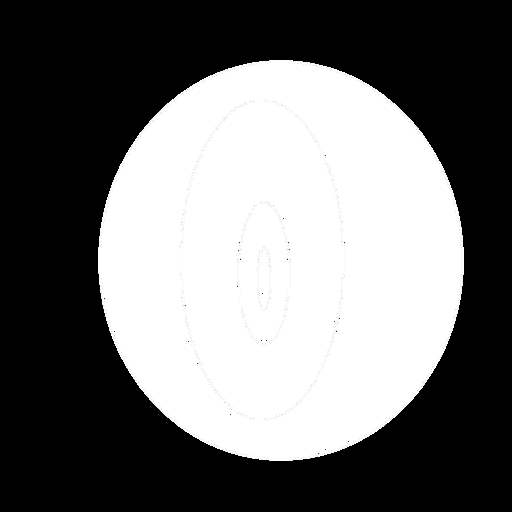}
    \label{fig:HoleIncorrect}
    }
    \quad
    \subfigure[Correct Fill]{%
    \includegraphics[width=3.75cm]{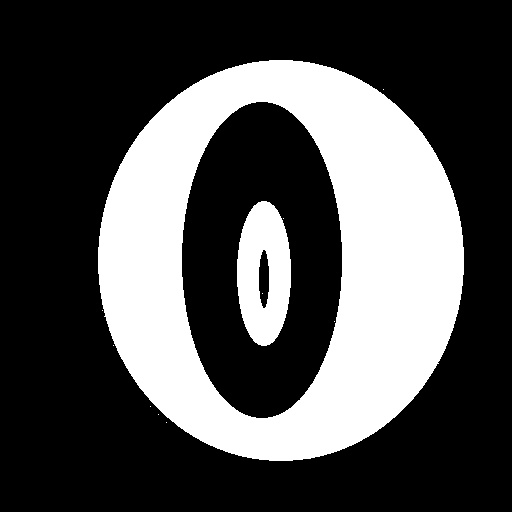}
    \label{fig:HoleCorrect}
    }
    \label{fig:GroupImage}
    \caption{A region filling example with `holes'. EFCI fails to fill the `holes' inside boundaries, since it cannot distinguish between adjacent connected regions such as 1 and 2, 2 and 3.}
\end{figure}

To handle the case with annuli, `holes' or even more complicated structures in regions, an improved version of the algorithm is needed. The intuition is that adjacent connected regions should be treated differently. For example, to fill an image with interior `holes', as described in Figure \ref{fig:HoleInput}, a desirable filling result should be Figure \ref{fig:HoleCorrect}, while EFCI tends to give the result in Figure \ref{fig:HoleIncorrect}. If we consider different connected areas to be either background or region to be filled, then adjacent connected regions should be of different types. To be precise, {\bf adjacent connected regions} refer to two connected regions $\mathbf{R_1}, \mathbf{R_2}$ such that:

\begin{equation}
\exists P_1 \in \mathbf{R_1}, P_2 \in \mathbf{R_2}: \overline{P_1 P_2} \subseteq \mathbf{R_1}\cup \mathbf{R_2} \cup \mathbf{B},
\end{equation}

where $\overline{P_1 P_2}$ denotes the line segment between $\mathbf{R_1}$ and $\mathbf{R_2},$ and $\mathbf{B}$ denotes the set of all boundary pixels of the regions in the image.

Intuitively, adjacent connected regions can be merged into one connected region if we take away all boundary pixels. In our example, the connected regions are labelled as 1,2,3,4,5 in Figure \ref{fig:HoleRegion}. 1 and 2, 2 and 3 are adjacent connected regions, hence should be treated differently.

To achieve the distinction between different connected regions, we need to introduce the classification of pixels.

\subsection{Classification of Pixels}
In the proposed algorithm, there are five types of pixels.

\noindent{\bf Type 1: Boundary pixels}

Boundary pixels are in {\bf boundary color} (pixel value 255 in our experiment), and are used to distinguish between different connected regions, especially during the process of flood filling. Their color is not changed during the implementation of the algorithm.

\noindent{\bf Type 2: Background pixels}

Background pixels are in {\bf background color} (pixel value 0 in our experiment), whose concept varies during the implementation.

In the {\bf Main Filling Process} of our algorithm (line 4 - 15) background pixels correspond to those pixels that have not yet been visited (filled, by either exterior or interior label color). Hence, when visiting a background pixel, the algorithm sets it as the seed of the flood filling algorithm, and then fill its connected region with either exterior or interior label color.

During the {\bf Crop-and-`Inverse' process} (line 16 - 23), exterior label pixels are set to the background color, since their label color merely represents temporary labels, instead of their final color.

\noindent{\bf Type 3: Exterior Label pixels}

Exterior label pixels are in {\bf exterior label color} (pixel value 80 in our experiment). In the Main Filling Process, they are labeled as future `background pixels', which are to be set back to background color during the Crop-and-`Inverse' process. However, they are colored in a temporary different color from the background, so that the algorithm can tell that they have been visited. A region whose pixels are in exterior label color is also useful in determining what color its adjacent connected region should be. In this case, its adjacent connected region should be in interior label color.

\noindent{\bf Type 4: Interior Label pixels}

Interior label pixels are in {\bf interior label color} (pixel value 128 in our experiment). In the Main Filling Process, they are labelled as future `mask pixels'. However, they are colored in a temporary different color from mask color, which is usually the same as boundary color, so that the algorithm can tell that they are not boundary pixels. A region whose pixels are in interior label color is also useful in determining what color its adjacent connected region should be. In this case, its adjacent connected region should be in exterior label color. Finally, they are set to mask color during the Crop-and-`Inverse' process.

\noindent{\bf Type 5: Mask pixels}

Mask pixels are in {\bf mask color} (pixel value 255 in our experiment), which do not appear until the  {\bf Crop-and-`Inverse' process}, when interior label pixels are set to mask color. Those mask pixels mutually make up the resulting masks for the output.

\begin{figure*}[htbp]
    \begin{center}
        \includegraphics[width=1\textwidth]{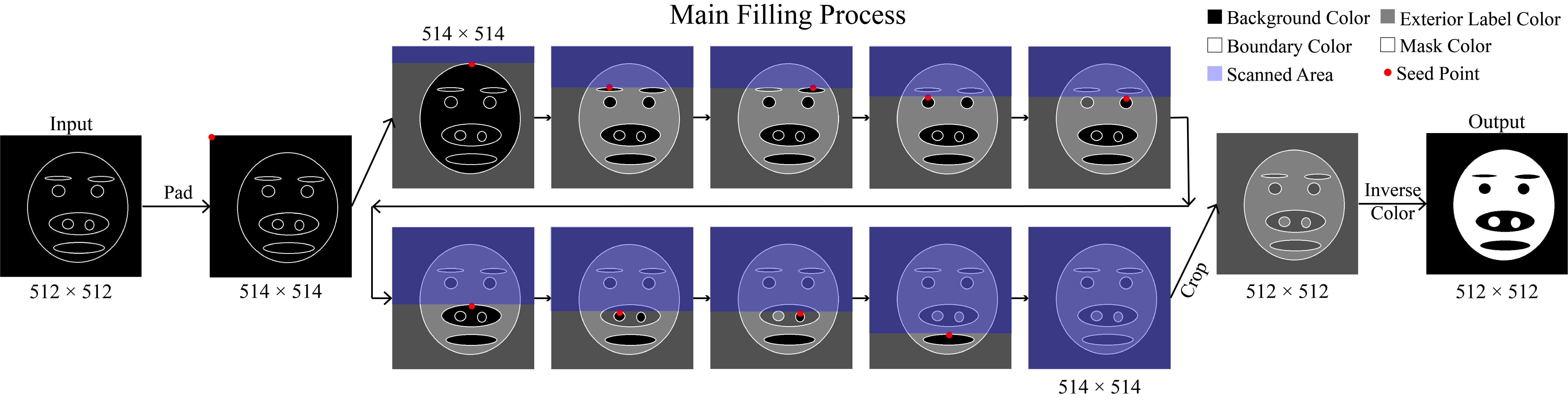}
    \end{center}
    \caption{Evolution of an image during the the implementation of Scan-flood Fill(SCAFF) algorithm. The flow chart illustrates how Scan-flood Fill(SCAFF) acts on an image: first pads the image, then labels different connected regions. After cropping, the algorithm sets different regions to resulting colors.The Main Filling Process is illustrated in details, where the algorithm scans through the image, searching for the next seed to start flood fill. The red pixel represents the next seed, and the blue area represents the scanned area. }
    \label{fig:ProcedureAdvance}
\end{figure*}

\subsection{Proposed Algorithm: Scan-flood Fill(SCAFF)}
We illustrate our algorithm below. Given an image with boundary pixels, padding is added. Then we start from the origin of the padded image and use flood filling algorithm to fill the connected region of the origin with label color. After that, we search through all other pixels in the image, setting each unprocessed non-boundary pixel (\ie a background pixel) as a new seed for flood filling, and fill its connected region with a certain color, either exterior or interior label color, based on the color of its adjacent filled connected region. In other to capture the color of its adjacent filled connected region, the algorithm searches backward until reaching a filled pixel, either an exterior or interior label pixel. When all background pixels are filled, we extract the image of the original size from the padded one and transform label color into the background color. Since our algorithm scans through the image to set potential starting seeds for flood filling, and scans backward to determine the color to be filled for the present visited region, we call it {\bf Scan-flood Fill(SCAFF).} An example of how the image changes is given in Figure \ref{fig:ProcedureAdvance}, and the pseudocode is given in algorithm \ref{algo:advance}.

\begin{algorithm}[!h]
    \KwIn{img}
    \KwOut{resultImg}
    padImg $\gets$ pad img with background color\;
    seed $\gets$ (0,0)\;
    padImg $\gets$ floodfill(padImg, seed) with label color\;
    \For{x in range(padImg.height)}{  \tcp*{Main Filling Process}
        \For{y in range(padImg.width)}{
            \If{padImg[x][y] == background color}{
                seed $\gets$ (x,y), i $\gets$ 1\;
                \While{padImg[x][y-i] == boundary color}{
                    i $\gets$ i + 1\;
                }
                \If{padImg[x][y-i] == boundary color}{
                    floodfill(padImg, seed) with filling color\;
                }
                
            }
            \If{resultImg[x][y] == exterior label color}{
                resultImg[x][y] $\gets$ background color\;
            }
            \Else{
                floodfill(padImg, seed) with label color\;
            }
        }
    }
    croppedImg $\gets$ crop padImg to original size, delete padding\; \tcp*{Crop-and-`Inverse' process}
    resultImg $\gets$ croppedImg\;
    \For{x in range(resultImg.height)}{
        \For{y in range(resultImg.width)}{
            \If{resultImg[x][y] == exterior label color}{
                resultImg[x][y] $\gets$ background color\;
            }
            \If{resultImg[x][y] == interior label color}{
                resultImg[x][y] $\gets$ mask color\;
            }
        }
    }
    \Return resultImg
    \caption{Scan-flood Fill(img)}
    \label{algo:advance}
\end{algorithm}

As illustrated above, this improved version of the region filling algorithm, Scan-flood Fill algorithm, can successfully deal with `holes', or even more complicated regions.

\section{Experiments}

We test both the basic version and our proposed algorithm on a set of images of different sizes, where regions inside have different properties. Comparing their resulting images, we also compute and analyze their running time. All experiments are performed on a PC with Intel(R) Core i7-8550U processor and 16GB RAM.\footnote {Codes and more related details are given in the following website: \url{https://github.com/SherylHYX/Scan-flood-Fill.}}

\begin{figure*}[htbp]
    \begin{center}
        \includegraphics[width=1\textwidth]{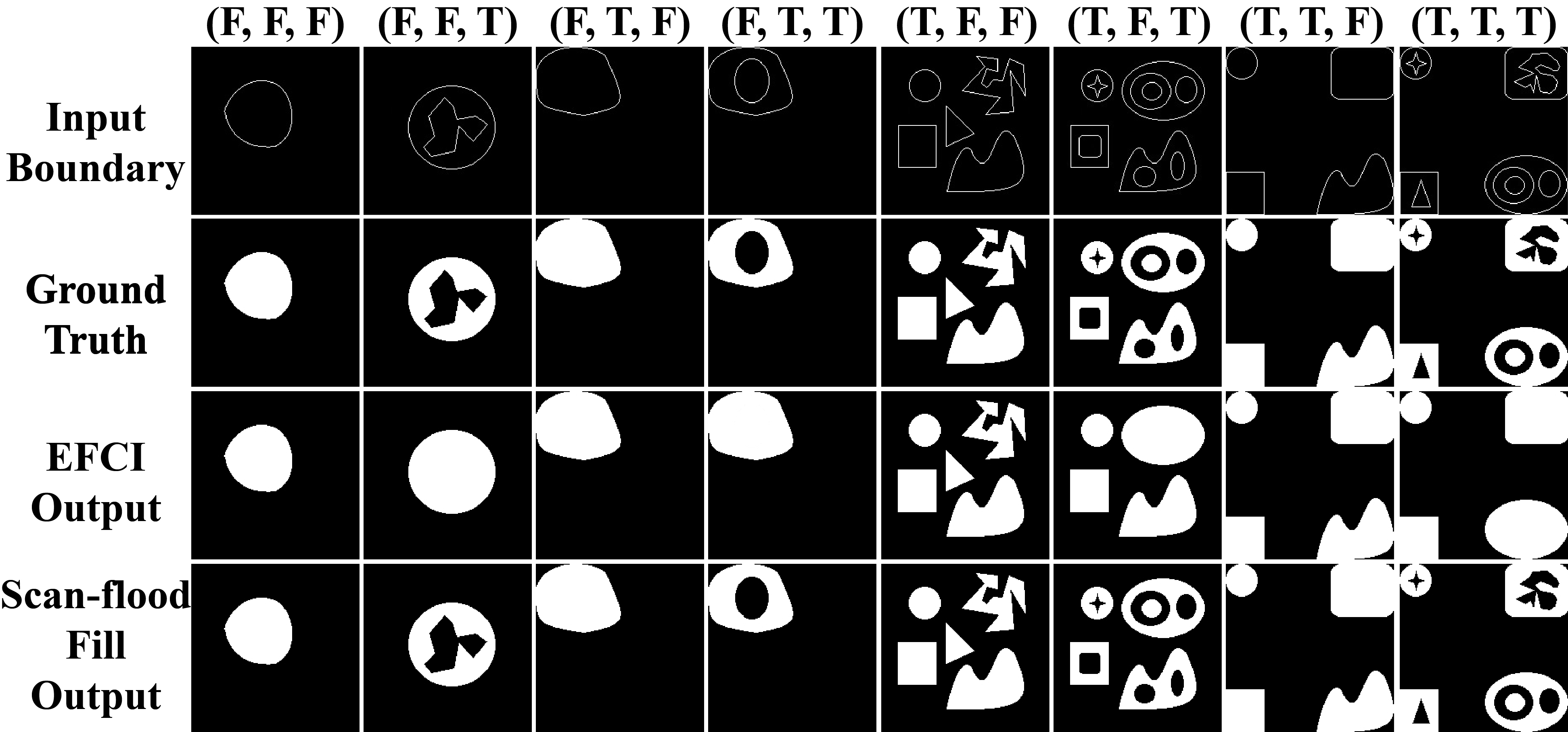}
    \end{center}
    \caption{Performance comparison on toy examples}
    \label{fig:Comparison}
\end{figure*}

\subsection{Datasets}

The input images for our toy experiment are generated by dilation of eight $200\times 200$ basic images. Each of the basic images represents a case in Table \ref{table:case}. Define the property 3-tuple of an image to be $(multiple, border, holes),$ whose entries are all Boolean variables, to represent each of these eight cases. For example, $(F,F,T)$ means that the image has only one object inside, has no regions whose boundaries meet the border of the whole image but has `holes' inside regions. This notation will be used to illustrate the comparison on toy examples. To show that our algorithm can also be useful in practical ground truth masks generating process, we also consider images in MSRA10K \cite{5}.

\subsection{Performance Comparison on Toy Examples}
For one thing, most filling algorithms require human-provided starting seeds. For another, among those automatic (with respect to seed selection) region filling algorithms, there are few that share the same filling purpose with us (\ie to fill the regions and to generate masks for machine learning). Hence, we may have different definitions for `accuracy'. For these reasons, we conduct the performance comparison and evaluation mainly on our proposed algorithm as well as its basic version. We first compare their performance on our toy examples. A visual result comparison is shown in Figure  \ref{fig:Comparison}.

From the result, we conclude that free from requiring users to provide initial seeds to start from, both algorithms can fill an arbitrary number of irregular regions without `holes' successfully. Thanks to the padding in the first step of each version of the algorithm, the result will not be influenced by regions whose boundaries lie on the boundary of the whole image. However, when it comes to more complicated regions such as annuli, rings and regions with interior `holes', only {\bf Scan-flood Fill} can generate desirable results.

\subsection{Time complexity evaluation}

Table \ref{table:time} illustrates the average running time of each image in folders with different sizes of input images. For a fair comparison, the time efficiency evaluations of both versions of the algorithm are performed on the same PC. Though Scan-flood Fill takes a bit longer time, the time consumed is not much longer, and the result is more precise.

\begin{table}
 \caption{The time(in seconds) consumed on different image sizes.}
\begin{tabular}{|l|l|l|}
\hline
\multirow{2}{*}{Size} & \multicolumn{2}{l|}{Version of Algorithm}   \\ \cline{2-3} 
                      & EFCI (s)   & Scan-flood Fill(SCAFF) (s) \\ \hline
200 $\times$ 200      & 0.07766529 & 0.1186108           \\ \hline
400 $\times$ 400      & 0.26739229 & 0.4606119           \\ \hline
600 $\times$ 600      & 0.58787846 & 1.0168558           \\ \hline
800 $\times$ 800      & 1.07662390 & 1.7586195           \\ \hline
1000 $\times$ 1000    & 1.60052500 & 2.7622317           \\ \hline
1200 $\times$ 1200    & 2.32550920 & 4.1238358           \\ \hline
1400 $\times$ 1400    & 3.16818610 & 5.6745079           \\ \hline
1600 $\times$ 1600    & 4.44347420 & 7.4507671           \\ \hline
1800 $\times$ 1800    & 5.45135353 & 9.4607772           \\ \hline
2000 $\times$ 2000    & 6.62144210 & 11.7155405          \\ \hline
\end{tabular}
\label{table:time}
\end{table}

\begin{figure}[htbp] 
    \centering
        \includegraphics[width=1\columnwidth]{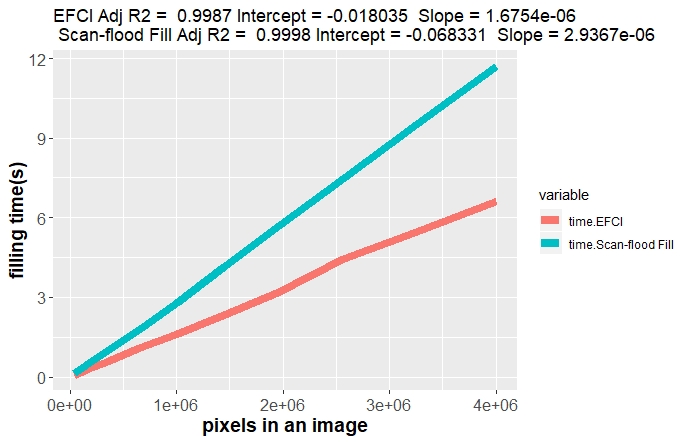}
    \caption{Running time comparison on toy examples.The time consumed is approximately proportional to the number of pixels in the image for each of EFCI and Scan-flood Fill. `Adj R2' means adjusted R square, `intercept' means the vertical intercept, and `slope' means the slope of the regression line.}
    \label{fig:Time_compare}
\end{figure}

We also analyze and compare the time complexity of each version of the algorithm. Plotting the number of pixels in an image and the time consumed, we obtain almost straight lines, as shown in Figure \ref{fig:Time_compare}. By regression, we claim that they both consume almost {\bf linear} time with respect to the number of pixels of an image, and the approximate complexities are both $\mathcal{O}(p),$ where $p$ denotes the number of pixels in the image. Furthermore, in order to save more time for Scan-flood Fill, it is also possible to consider cropping immediately after the first flood fill in the Main Filling Process, \ie exactly after line 3 in Algorithm \ref{algo:advance}, and then scan through the cropped image, instead of cropping after the whole Main Filling Process.

\subsection{Practical Use to Generate Ground Truth Masks}
\begin{figure*}[htbp]
    \begin{center}
        \includegraphics[width=1\textwidth]{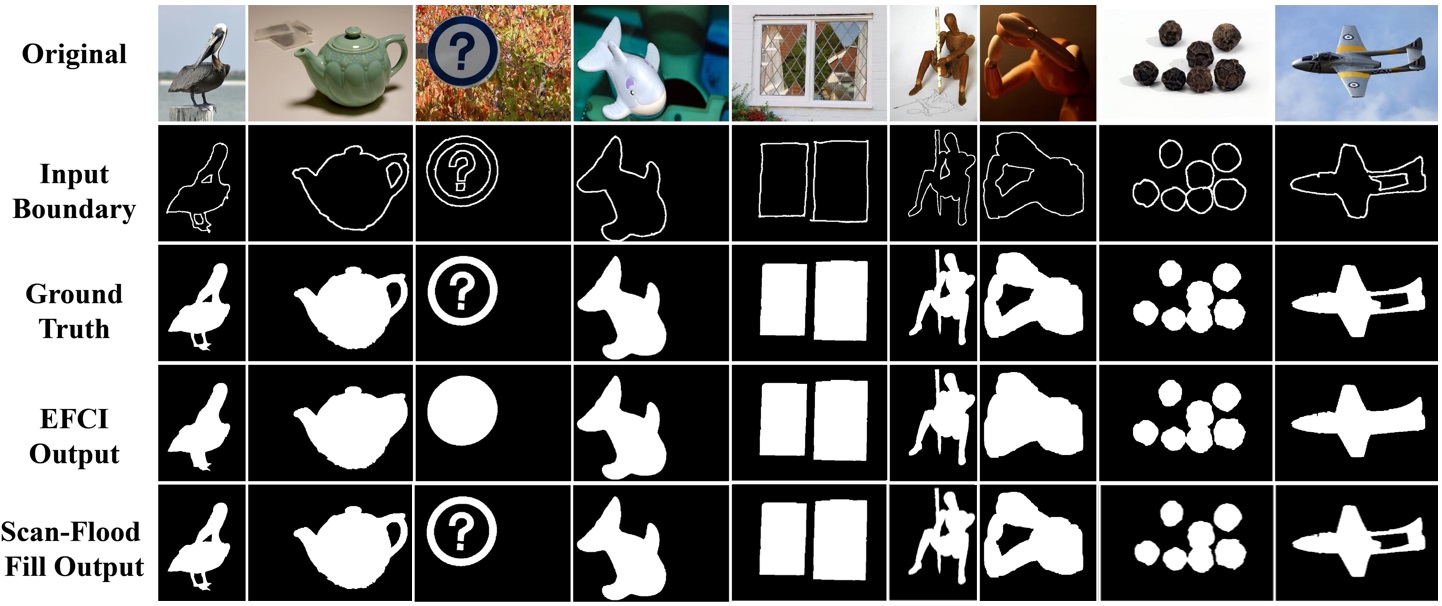}
   \end{center}
    \caption{Some practical examples to generate ground truth masks for supervised learning, \ie, from left to right: different images with different types of object shapes; from top to bottom: original input images, boundary images, ground truth images, the corresponding output images with EFCI, and the corresponding output images with Scan-flood Fill(SCAFF).The Scan-flood Fill(SCAFF) results are almost the same as ground truth images, so they can be used as ground truth images and would probably not affect training accuracy.}
    \label{fig:Practice}
\end{figure*}

We test Scan-flood Fill Algorithm as well as its basic version(EFCI) on 9,918 out of 10,000 images from MSRA10K \cite{5}, where edges are relatively easy to be extracted from masks(since this dataset does not provide edge information). Starting from an image with `only' boundary pixels (\ie masks not yet generated), Scan-flood Fill can generate corresponding masks effectively. To be comparable with ground truth masks given by MSRA10K \cite{5}, we set mask color to be boundary color, \ie, pixel value 255 in our case. A visual result for some of the images is shown in Figure \ref{fig:Practice}. 

We also obtain a comparison of quantitative results including F1 score\cite{goutte2005probabilistic} (larger is better) and MAE\cite{coyle1988stack}(Mean Absolute Error, smaller is better), as is given in Table \ref{table:evaluation}. The quantitative results indicate that Scan-flood Fill achieves better performance than EFCI on the given dataset. The difference between them probably lies in the existence of `holes' within regions to be filled. Besides, since MSRA10K \cite{5} does not provide edge images, and the generating process of edges may result in differences of boundary pixels between ground truth images and generated edge images, it is reasonable that Scan-flood Fill cannot achieve 100 percent accuracy in this case, when compared to GT masks given by the dataset. The Scan-flood Fill results are almost the same as ground truth results, so they can be used as ground truth for supervised learning and would probably not affect training accuracy.

\begin{table}
 \caption{Comparison of quantitative results including F1 score (larger is better) and MAE(Mean Absolute Error, smaller is better), on dataset MSRA10K.}
\begin{tabular}{|l|l|l|}
\hline
\multirow{2}{*}{Metric} & \multicolumn{2}{l|}{Version of Algorithm}   \\ \cline{2-3} 
                      & EFCI    & Scan-flood Fill  \\ \hline
F1 score     & 0.985600323 &  0.988181049           \\ \hline
MAE      & 0.005224898 &  0.004115951           \\ \hline
\end{tabular}
\label{table:evaluation}
\end{table}

\subsection{Advantages of Scan-flood Fill(SCAFF) Algorithm}
In previous works, seed filling algorithms such as flood filling algorithm from OpenCV and boundary filling algorithms have been applied to generate filled masks \cite{bradski2008learning,howse2013opencv}. Although these are used in part of our approach, there exists a considerable difference in that our {\bf starting seeds are automatically provided} by the algorithm instead of being given beforehand. We propose to integrate seed filling algorithms and scan-line filling algorithms, together with the properties of adjacent connected regions, for complicated arbitrary region filling problems. 

Besides, Scan-flood Fill(SCAFF) enables us to handle complicated regions such as a pig face and the examples from MSRA10K \cite{5}. This is by virtue of the classification of pixels, especially label pixels, and the Crop-and-`Inverse' process. We compare Scan-flood Fill(SCAFF) with its basic version EFCI for region filling. The results in Figure \ref{fig:Comparison} demonstrates the potential superiority of more precise algorithms taking the relationship between adjacent connected regions into account.

Also, in Scan-flood Fill(SCAFF) algorithm, we do not need to worry about the potential existence of {\bf multiple regions inside an image}. This is because our filling algorithm does not require human-provided starting seeds, but can set starting seeds whenever needed. Moreover, since we start from filling exterior label color to the outermost exterior of an image, which is part of the background for the final result, we are able to avoid being trapped in a small region. 

Moreover, padding with background color guarantees the robust filling result regardless of whether some boundaries of the regions in an image lie on the border of the whole image. 

\section{Conclusion}
In this paper, we present an efficient automatic region filling algorithm for arbitrary regions by using a color scheme to assign different labels to adjacent connected regions. The algorithm scans through all pixels in the image, automatically sets the seeds for flood filling, and labels all visited pixels, before the Crop-and-`Inverse' process. The resulting masks distinguish pixels intended to be filled from background efficiently. The Scan-flood Fill(SCAFF) algorithm is effective in generating masks as ground truth images used for supervised model training. Experiments on various types of images clearly demonstrate the effectiveness of our algorithm, and robustness when handling multiple regions, complicated `holes' and regions intersecting image border. Besides, as the problem confronted is to judge the interior pixels bounded by some arbitrary shapes, we still consider the proposed algorithm to be eligible in giving some possible potential insight into overcoming the related graphics or computational geometry problems.   

\section{Acknowledgement}
This work was supported in part by grants from National Science Foundation of China (No.61571005, No.61811530271), the China Scholarship Council (CSC NO.201806155037), the Science and Technology Research Program of Guangzhou, China (No.201804010429), the Fundamental Research Funds for the Central Universities, SCUT (No.2018MS57).

{\small
\bibliographystyle{ieeetr}
\bibliography{egbib}
}

\end{document}